\documentclass[aps,prb,preprint,12pt]{revtex4-1}
\usepackage{amssymb,euscript,latexsym,amsmath}
\usepackage{graphicx}
\usepackage{color}
\usepackage{tikz,pgfplots}
\usepackage{xcolor}
\usepackage{amsmath}
\usepackage{tikz}
\usepackage{hyperref}
\usepackage[makeroom]{cancel}
\begin{document}
%\tdplotsetmaincoords{70}{110}
\title{Coulomb force between two Dirac monopoles} 
%\author{David Garfinkle}
\author{Alberto G. Rojo}
%\email{garfinkl@oakland.edu}
\affiliation{Department of Physics, Oakland University, Rochester, MI 48309}
\email{rojo@oakland.edu}
%\affiliation{Department of Physics, Oakland University, Rochester, MI 48309}

%\received{June 18, 2013}
%\accepted{}

%Editors Note. {The concept of magnetic monopoles was proposed almost a century ago by Paul Dirac.  While no such entity has been detected, it remains a powerful theoretical concept as the existence of even a single magnetic monopole in the Universe would explain the quantization of electric charge. To avoid violating Maxwell’s equations, Dirac’s model involved the concept of an infinitely thin string-like solenoid. This paper shows that such strings will interact with each other with an inverse-square Coulomb-like force; this is done by modeling the monopoles as semi-infinite lines of magnetic dipoles.  This analysis would be appropriate for advanced students of electromagnetism.}

\begin{abstract}

The model of magnetic monopoles that was proposed by 
 Paul Dirac in 1931 has long been a subject of theoretical interest in physics because of its potential to explain the quantization of electric charge. While much attention has been given to non-Dirac monopoles, Dirac's model, which involves an infinitely thin solenoid known as a Dirac string, presents subtleties in the interaction between monopoles. In this paper, we show that the force between two Dirac monopoles  obeys a Coulomb-like interaction law. This derivation offers an instructive exercise in fundamental electromagnetism concepts and is appropriate for undergraduate and early graduate-level students. 

\end{abstract}

\maketitle

\section{Introduction}

The concept of magnetic monopoles has intrigued physicists since  Paul Dirac proposed their existence in 1931.\cite{Dirac} Dirac showed that if even a single magnetic monopole existed in the universe, it would explain the quantization of electric charge, a fundamental empirical fact. 
Dirac's monopole model  introduces  an infinitely thin string, now referred to as a Dirac string, which enables the monopole to exist without violating Maxwell's equations.\cite{Jackson}
Although no magnetic monopole has been detected experimentally, they remain an important theoretical construct,
playing roles in areas such as quantum field theory, grand unified theories,  gauge symmetries, and other areas of theoretical physics. \cite{Siva, unified, otherA,PTODAY, Heras,Franklin, Preskil, gauge}

Although the Coulomb interaction between monopoles is relatively straightforward in the case of non-Dirac monopoles where \(\nabla \cdot \mathbf{B} =\mu_0  g \delta(\mathbf{x})\) with $g$ being the magnetic charge, it is less intuitive that  Dirac monopoles would follow the same Coulomb-like interaction.\cite{Griffiths1} In this paper we present a  derivation of this result, showing that Dirac monopoles also obey a Coulomb-like force law despite the presence of the Dirac string. This calculation can serve as a useful exercise in electromagnetism classes, offering students insight into the behavior of monopoles in both conceptual and computational contexts.

In the following sections, we first derive the field of a Dirac string by modeling the monopole as a semi-infinite line of dipoles.  This derivation offers a simpler alternative to the usual calculation, which is based on the vector potential for a straight string.\cite{Heras}    {\color{black} By subsequently computing the interaction between strings through a direct force method, we show that the static Coulomb force between Dirac monopoles emerges in a natural way.}

\section{Field of a Dirac string }
\begin{figure}[h]
\begin{tikzpicture}[scale=2]
%
%     \draw[line width=1.2,->]  (0,0)  -- (1.48,-1.65)  node[midway, sloped, above] {$\mathbf  x$};
% Eliptical dirac string
  \draw[fill=gray!40,draw=none] (0,-2.5) ellipse (0.035 and 0.01);
  % Cylinder body (extra thin)
  \shade[left color=gray!20, right color=gray!80,]
    (-0.035,-.02) rectangle (0.035,-2.5);

  % Elliptical caps (extra thin)
 % top
 % \draw (0,-.05) ellipse (0.035 and 0.01);               % bottom outline

  % Label (optional)
 % \node[right] at (0.15,1.5) {Dirac string};
 %-
    % Draw the central point
    \filldraw (0,0) circle (1pt); % The center of the diagram

    % Draw the arrows radiating from the center
     \begin{scope}[rotate=22.5]
    \draw[thick,->] (0,0) -- (1,0);  % Right
   \draw[thick,->] (0,0) -- (-1,0); % Left
    \draw[thick,->] (0,0) -- (0,1);  % Up
     \draw[thick,->] (0,0) -- (0,-1);  % Up
%   \draw[thick,->] (0,0) -- (0,-1); % Down
    \draw[thick,->] (0,0) -- (0.7,0.7);  % Diagonal up-right
 \draw[thick,->](0,0) -- (-0.7,0.7); % Diagonal up-left
 \draw[thick,->] (0,0) -- (-0.7,-0.7);% Diagonal down-left
   \draw[thick,->] (0,0) -- (0.7,-0.7); % Diagonal down-right
\end{scope}
    % Label for "g"
    \node at (0.,0.3) {$g$};

    % Label for "B"
    \node at (1.,0.6) {$\mathbf{B}_{\rm mon}$};

    % Draw the dashed curved line (particle trajectory)

%\draw[very thick, gray] (0,-2)--(0.,0);
    
%    \draw[very thick, gray] plot [smooth, tension=1] coordinates {(0,0) (0.1,-1) (-0.5,-2.5) (-0.2,-3.5)};
% \draw[thick, gray,->] (-0.36,-2.0) -- (-.31,-1.9) node[left] {$\Phi$};
 %
 %
  \draw[line width=2.5,->] (0.,-1.4) -- (0,-1.2) node[midway,left] {$d \mathbf m={g}dz \hat{\mathbf{k}}$};
 %
%----------------------------------------------
%
%
 %
 \begin{scope}[shift={(.49,1.1)}]
 %\draw[line width=1.1,->]  (-0.475,-2.43)  -- (1,-2.8) ;
 %node[right] {$\mathbf {x}$};

 %
 %\node  at (0.25,-2.5)[rotate=-14.5]{$\mathbf {x}-\mathbf {x}'$};
 %
  %\draw[fill] (1.02,-2.8) circle(0.02);
 \end{scope}
 %\draw (-0.04,-1.6) --(-.8,-1.9) node[left]{$\gamma$};
 \draw (-0.05,-1.6) .. controls (-0.25,-1.66) .. (-0.5,-1.9)
      node[left]{$\gamma$};

\end{tikzpicture}
\caption{Dirac string consisting of a one dimensional semi-infinite line $\gamma$ 
along the $z$-axis that terminates at the origin and carries
a magnetic flux $\Phi$. The field at any point is the superposition of the fields created by infinitesimal dipoles $d\mathbf m$ that point along the direction of  $\gamma$. 
At any point outside the string, the field is that of a monopole,
$\mathbf{B}_{\rm mon}$,  of magnetic charge $g=\Phi/\mu_0$, which emanates in all directions from the end of the string.  }
\label{String}
\end{figure}
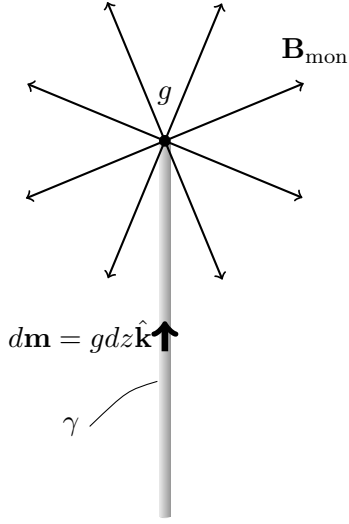

Following Dirac, we consider  a  magnetic monopole as a semi-infinite, infinitesimally thin solenoid described by the line (or string) $\gamma$ (see Figure \ref{String}).
 %%%%%%%%%%%
%%%%%%%%%%%%%%%
 If  $I$ is the current in the solenoid, the field inside is $B=\mu_0 I n$, with $n$ being the number of turns per unit length. In addition, the magnetic moment of each turn is $m=I A$, with $A$ being the cross-sectional area of the solenoid.  This means that the field of an infinitely thin solenoid can be regarded as a continuous semi-infinite line $\gamma$ of magnetic dipoles. The flux through the solenoid is $\Phi= BA= \mu_0InA$.\cite{Clarification} For a Dirac string, $A\rightarrow 0$ and $In\rightarrow\infty$ in such a way that the product $InA=g$  remains finite,    with $g$  being a magnetic moment per unit length:
\begin{equation}
g= {\Phi \over \mu _0}.
\end{equation} 

{\color{black} 
This   semi-infinite  line of dipoles extending from $z\rightarrow -\infty$  to the origin corresponds to a magnetization (or magnetic moment per unit volume) $
\mathbf M(\mathbf x)$  given by
\begin{equation}
\mathbf M(\mathbf x)= g\hat{\mathbf k} \delta(x) \delta(y) \theta (-z),
\label{magnet}
\end{equation}
with $\theta(z)$ the Heaviside function.
Since the magnetic field $\mathbf B (\mathbf x)$  in all space is given by\cite{Griffiths2}
\begin{equation}
\mathbf B(\mathbf x)= \mu_0\left(\mathbf M(\mathbf x) +\mathbf H(\mathbf x)\right),
\end{equation}
and $\nabla \cdot \mathbf B=0$, we have
\begin{equation}
\nabla \cdot \mathbf H(\mathbf x)= -\nabla \cdot \mathbf M(\mathbf x)=g\delta(\mathbf x),
\end{equation}
where we used Eq. (\ref{magnet}) and the fact that $\theta'(z)=\delta(z)$. In other words, the source of $\mathbf H$ is a point charge $g$ at the origin, giving \begin{equation}\mathbf H(\mathbf x) ={g\over 4\pi } {\mathbf x\over |\mathbf x|^3}.\end{equation}
The full magnetic field of the string is therefore
\begin{equation}
\mathbf{B}_\gamma(\mathbf x)=\frac{\mu_0}{4 \pi} g {\mathbf x\over |\mathbf x|^3
} +\mu_0 g \delta(x) \delta(y) \theta(-z) \hat{\mathbf{z}}.
\label{bstring}
\end{equation}

We observe that the flux through a surface enclosing the origin is zero, since it has contributions $+\mu_0g $ from the the monopolar field of the first term in Eq. (\ref{bstring})   and $-\mu_0g$
from the second term (the string), as expected for a field with zero divergence.}

\section{Coulomb force between Dirac strings}

The present  evaluation of the Coulomb force is  based on the fact that
the force on a point dipole {\color{black} defined by the vector}  $\mathbf m$ 
{\color{black} located at the position $\mathbf x$ in the presence of a magnetic field $\mathbf {B} (\mathbf{x})$} is\cite{boyer}
\begin{equation}
\mathbf{F}=\nabla \left(\mathbf m\cdot \mathbf B(\mathbf x) \right).
\label{forceondipole}
\end{equation}

%{\color{red} Eq. (\ref{forceondipole}) derives from the }

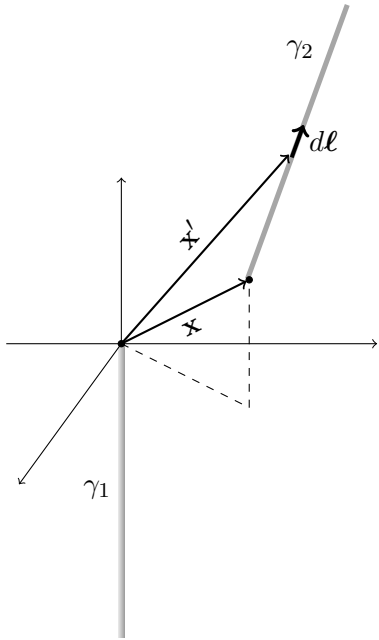
\begin{figure}[h]
\begin{tikzpicture}[scale=1.7]
  \shade[left color=gray!20, right color=gray!80,]
    (-0.025,.02) rectangle (0.025,-2.3) node[midway,left] {$\gamma_1$};
%\draw [very thick,->](0,0)--node[midway,left] {$\gamma_1$}(0,1.5);
%\node[above] at (0,1.5) {$z$};
\draw [->](-.9,0)--(2,0);
\draw [<-](0,1.3)--(0,0);
\draw [->] (0,0)--(-.8,-1.1);
\draw [dashed] (0,0)--(1,-0.5);
\draw [dashed] ((1,-0.5)--(1.,.5);
\draw [thick, ->] (0,0)--(.98,.49) node[midway,sloped,below]  {$\mathbf x$};
%\draw [blue,very thick,->] (1,.5)--(1.9,1.5) 
%node[midway,above left] {$\gamma_2$}
;
\begin{scope}[shift={(.98,.49)}]
\begin{scope}[rotate=-20]
  \fill[color=gray!70,]
    (-0.02,.02) rectangle (0.02,2.3);
      \draw[line width=1.5,->] (0.,1.03) -- (0,1.3) node[midway,right] {$d\boldsymbol{\ell}$};
\end{scope}
\end{scope}

\node at (1.4,2.3) {$\gamma_2$};
%\node[above] at (.91,.49) {$\mathbf x$};
%\node[above] at (2.05,1.35) {$\mathbf x'$};
   \filldraw (0,0) circle (.7pt);
      \filldraw (1,.5) circle (.7pt);
      \draw [thick, ->] (0,0)--(1.31,1.48) node[midway,sloped,above]  {  $ \mathbf x'$};
\end{tikzpicture}
\caption{Two Dirac strings $\gamma_1$ and $\gamma _2$ with end points at the origin and at $\mathbf x$ respectively.}
\label{Twostrings}
\end{figure}

Suppose that we have a Dirac string \(\gamma_1\) terminating at the origin and a second string \(\gamma_2\) terminating at the point \(\mathbf{x}\) as shown in Figure \ref{Twostrings}.

The field $\mathbf B_1$ of the string $\gamma_1$ at the location of $\gamma_2$ is that of a monopole of charge $g_1$ at the origin; the second term in Eq. (\ref{bstring}) is zero everywhere except on the string $\gamma_1$.
 At the same time, as derived in the previous section, the string \(\gamma_2\) can be regarded as a continuous line of dipoles with a dipole moment per unit length given by \(g_2\). Note that if $g_2$ is to be positive (a ``north" pole), the dipoles must point inward: $d \mathbf{m}=-g_2 d \boldsymbol{\ell}$ . 
The force on each element of the string $\gamma_2$ is thus $$d\mathbf F = \nabla \left( -g_2 d\boldsymbol {\ell} \cdot {\mu_0 g_1 \over 4\pi} {\mathbf x' \over  |\mathbf {x} '|^3} \right),$$ and the total force is found by integrating this over the length of the string. Since the gradient is with respect to a different variable, $\mathbf {x}'$, the order of integration and differentiation can be interchanged, yielding:
\begin{eqnarray}
\mathbf{F}_{2,1} & =&\boldsymbol{\nabla} \int_{\gamma_2} d \mathbf{m} \cdot \mathbf{B}_1=-\frac{\mu_0}{4 \pi} g_1 g_2 \boldsymbol{\nabla} \int_{\gamma_2} d \boldsymbol{\ell} \cdot \frac{ \mathbf{x}'}{\left | \mathbf{x}'\right|^3}. 
%\\
%& =&-\frac{\mu_0}{4 \pi} g_1 g_2 \boldsymbol{\nabla} \int_x^{\infty} \frac{d x^{\prime}}{x^{\prime 2}}=-\frac{\mu_0}{4 \pi} g_1 g_2 \boldsymbol{\nabla} \frac{1}{x}=\frac{\mu_0}{4 \pi} g_1 g_2 \frac{\hat{\mathbf{x}}}{x^2},
%
\end{eqnarray}

{\color{black}

Compare this line integral to that which is used to calculate the potential difference between the two ends of a line $\gamma$ in the presence of a point charge $q$ located at  the origin
(note that we are reversing the normal usage of  $\mathbf{x}'$):

\begin{equation}
    \Delta V= -\int _\gamma \mathbf{E}(\mathbf{x}')\cdot d\mathbf{\boldsymbol \ell}=-{q\over 4\pi \epsilon_0} 
     \int _{\gamma} d\mathbf{\boldsymbol \ell} \cdot  {\mathbf {x}' \over |\mathbf{x'}|^3}.
\end{equation}

These two integrals have the same form. For the electrostatic problem, we know that potential difference is $\Delta V = \left({1/4\pi \epsilon_0}\right) (q/r_2 - q/r_1)$. For a semi-infinite line, $r_2\rightarrow \infty$ and  $r_1 = r$, the distance from the origin to the point $\mathbf{x}$, 
 $\Delta V =- {q/4\pi \epsilon_0}r$.
Using this result, we see:

\begin{eqnarray}
\mathbf{F}_{2,1} & =&-\frac{\mu_0}{4 \pi} g_1 g_2 \boldsymbol{\nabla} \frac{1}{r}=\frac{\mu_0}{4 \pi} g_1 g_2 \frac{\hat{\mathbf{r}}}{r^2},
\end{eqnarray}
a Coulomb-like interaction as claimed.}

\section{Acknowledgements}

We thank David Garfinkle,  David J. Griffiths,  Eduardo Jagla, Jorge Russo and Jorge Sofo for useful comments on the manuscript.


\begin{thebibliography}{2}
\bibitem{Clarification} 

Strictly speaking, the given expression corresponds to the flux and field of an infinite solenoid, from which we extract a magnetic moment per unit length. By simple symmetry considerations, the magnetic field at the end of a semi-infinite solenoid along its axis, is $\frac{1}{2} \mu_0 n$. However, this factor of $1/2$ affects only the local field at the solenoid’s end and does not alter the dipole density, thus having no impact on our calculations.
\end{thebibliography}
\end{document}